\documentclass[a4paper,fleqn]{cas-dc}

\usepackage[numbers,sort&compress]{natbib}
\usepackage{amssymb}
\usepackage{amsmath}
\usepackage{graphicx}
\usepackage{subcaption}
\usepackage{multirow}
\usepackage{algorithm2e}
\usepackage{booktabs}
\usepackage{tabularx}
\usepackage{array}
\usepackage{makecell}
\usepackage{xcolor}
\usepackage{url}
\usepackage{hyperref}
\usepackage{siunitx}
\usepackage{placeins}

\graphicspath{{figures/}}

\newcolumntype{Y}{>{\centering\arraybackslash}X}

\begin{document}

\title[mode=title]{Two-Stage Deformable-Convolutional Inverse Design of Nanophotonic Absorbers from Optical Spectra}
\shorttitle{Two-Stage Deformable Nanophotonic Inverse Design}
\shortauthors{Waseer et al.}

\author[1]{Waleed Waseer}
\ead{waleedwaseer@gmail.com}
\credit{Conceptualization of this study, Methodology, Investigation, Writing -- review and editing}

\author[2]{Muhammad Shahid Jabbar}[orcid=0000-0003-2331-876X]
\ead{muhammad.jabbar@kfupm.edu.sa}
\credit{Investigation, Formal Analysis, Writing -- original draft, Writing -- review and editing}

\author[3]{Muhammad Sohail Ibrahim}[orcid=0000-0002-1387-0879]
\ead{muhammad.ibrahim@kfupm.edu.sa}
\credit{Visualization, Writing -- original draft, Writing – review and editing}

\author[4,2]{Shujaat Khan}
\cormark[1]
\ead{shujaat.khan@kfupm.edu.sa}
\credit{Conceptualization of this study, Methodology, Supervision, Investigation, Formal Analysis, Writing -- original draft, Writing -- review and editing}
\cortext[1]{Corresponding author}

\affiliation[1]{organization={School of Physics and State Key Laboratory of Electronic Thin Films and Integrated Devices, University of Electronic Science and Technology of China},
    city={Chengdu},
    postcode={610054},
    country={China}}

\affiliation[2]{organization={SDAIA-KFUPM Joint Research Center for Artificial Intelligence, King Fahd University of Petroleum $\&$ Minerals},
    city={Dhahran},
    postcode={31261}, 
    country={Saudi Arabia}}

\affiliation[3]{organization={Interdisciplinary Research Center for Intelligent Secure Systems (IRC-ISS), King Fahd University of Petroleum $\&$ Minerals},
    city={Dhahran},
    postcode={31261}, 
    country={Saudi Arabia}}

\affiliation[4]{organization={Department of Computer Engineering, College of Computing and Mathematics, King Fahd University of Petroleum \& Minerals},
    city={Dhahran},
    postcode={31261},
    country={Saudi Arabia}}

\begin{abstract}
Data-driven inverse design provides an efficient means of generating nanophotonic structures with prescribed optical responses, but spectrum-to-geometry mapping is inherently non-unique and challenging for geometries containing thin elements, narrow gaps, and sharp spatial transitions. This work presents a two-stage deformable-convolutional framework for reconstructing metal--insulator--metal resonator geometries from absorption spectra. An 80-dimensional spectrum is projected to a $150\times4\times4$ spatial latent representation and decoded into a $64\times64$ resonator mask using deformable convolutional layers. Training is performed in two stages: supervised reconstruction establishes the global spectrum-to-geometry mapping, followed by least-squares adversarial refinement initialized from the best supervised checkpoint. A three-run ablation compares deformable convolution with plain convolution, involution, Dynamic Conv, and ODConv under the same decoder architecture. The proposed two-stage DeformConv model achieves $20.79\pm0.31$~dB PSNR and $0.8501\pm0.0082$ SSIM, outperforming plain convolution by 2.16~dB and 0.0831, respectively. Binary geometric evaluation yields a Dice score of $0.9623\pm0.0027$, IoU of $0.9342\pm0.0038$, boundary F-score of $0.9550\pm0.0027$, HD95 of $1.883\pm0.109$ pixels, and average surface distance of $0.353\pm0.024$ pixels. Spectral consistency is assessed by passing thresholded predictions through a separately trained frozen forward surrogate, yielding RMSE of $0.0805\pm0.0013$ and $R^2=0.7923\pm0.0065$. Analysis of learned deformable offsets reveals scale-dependent sampling, with stronger geometry-associated displacement at coarse and intermediate decoder stages than at final resolution. These results demonstrate that adaptive spatial sampling combined with supervised initialization and adversarial refinement improves spectrum-conditioned geometry reconstruction.
\end{abstract}



\maketitle

\section{Introduction}
\label{sec:introduction}

Nanophotonic structures manipulate electromagnetic fields through subwavelength geometry and have enabled compact components for sensing, imaging, spectral filtering, thermal emission, and wavefront control \cite{molesky2018inverse,yao2019intelligent,ma2021deep,jiang2021deep}. A common implementation is the metal--insulator--metal (MIM) absorber, where a patterned metallic resonator is separated from a metallic back reflector by a dielectric spacer. Small changes in the resonator outline, arm length, gap width, or symmetry can shift resonance locations and amplitudes. Consequently, the design problem is strongly nonlinear and high dimensional.

Conventional forward design evaluates a candidate geometry using an electromagnetic solver such as the finite-difference time-domain (FDTD) method and then modifies the geometry through parameter sweeps, topology optimization, evolutionary search, and adjoint optimization \cite{molesky2018inverse}. These approaches can produce high-performance devices, however, repeated full-wave simulations are expensive. Machine learning surrogates provide an alternative by learning the geometry-to-spectrum map from simulated data and evaluating new structures rapidly \cite{ma2021deep,jiang2021deep,yeung2020elucidating}. The more difficult inverse problem seeks a geometry for a specified spectrum. It is inherently non-unique, where multiple designs may produce similar responses, while a direct regression loss encourages the network to average across plausible solutions and can therefore produce blurred or physically ambiguous outputs \cite{liu2018training,cho2026probabilistic}.

Several neural strategies have been proposed to address this problem. Tandem networks pass the predicted design through a pretrained forward model and optimize spectral consistency rather than requiring a unique structural label \cite{liu2018training,xu2023inverse}. Generative adversarial networks (GANs) learn a distribution of realistic structures and have been used for metamaterial, metagrating, and metasurface design \cite{liu2018generative, jiang2019free, so2019designing, wen2020robust}. Hybrid frameworks combine generative initialization with conventional electromagnetic optimization \cite{yeung2022deepadjoint}. More recently, diffusion and probabilistic models have been introduced to improve solution diversity, physical consistency, and fabrication awareness \cite{hen2025inverse, seo2026physics, mondal2026mxdiffusion, cho2026probabilistic}. Despite this progress, a practical deterministic inverse model remains valuable when a paired design is available and fast one-shot reconstruction is required.

Another challenge pertains to the architectural design. Standard convolution samples a fixed Cartesian grid at every spatial location. Such stationarity is effective for natural-image texture but is not necessarily optimal for resonator masks containing variable arm thicknesses, narrow gaps, disconnected components, rounded corners, and inverted foreground/background patterns. Deformable convolution augments the regular sampling grid with learned offsets, allowing the receptive field to adapt to local geometry \cite{dai2017deformable, zhu2019deformable}. This capability motivates its use in the spectrum-to-shape decoder studied here.

To address these challenges, this paper proposes a two-stage deformable-convolutional inverse model for the open MIM nanophotonic dataset introduced by Yeung \emph{et al.} \cite{yeung2020elucidating}. The design choices are intentionally simple and controlled. First, a supervised generator learns the global spectrum-to-geometry mapping using a reconstruction objective. Second, the best supervised checkpoint initializes an adversarial stage based on least-squares GAN (LSGAN) loss \cite{mao2017least}, where the reconstruction term is retained to preserve correspondence with the paired target while the adversarial term refines fine structural details. To isolate the effect of spatial operator choice, the same network is trained with standard convolution, deformable convolution, involution \cite{li2021involution}, dynamic convolution \cite{chen2020dynamic}, and omni-dimensional dynamic convolution (ODConv) \cite{li2022omni}.

The main contributions are as follows:
\begin{itemize}
    \item We formulate spectrum-conditioned MIM absorber reconstruction as generation of a $64\times64$ resonator mask from an 80-dimensional absorption vector.
    \item We introduce a compact decoder with four deformable convolutional layers following a learned spectral projection, enabling spatially adaptive sampling during geometry formation.
    \item We employ supervised MSE pretraining followed by initialized LSGAN refinement and compare this strategy with both supervised-only and single-stage adversarial training.
    \item We conduct a three-run controlled operator ablation covering plain convolution, deformable convolution, involution, dynamic convolution, and ODConv.
    \item We evaluate the proposed model with different evaluation metrics and repeated the analysis for three independent runs.
    \item We perform spectral round-trip validation by passing inverse predictions through an independently trained forward surrogate and measuring global and resonance-peak agreement.
\end{itemize}

\section{Background}
\label{sec:related}
\subsection{Learning-based Nanophotonic Inverse Design}

Early data-driven inverse design methods primarily targeted low-dimensional structural parameters. Liu \emph{et al.} introduced a tandem architecture that connects an inverse network to a pretrained forward network, mitigating conflicting structural labels by measuring error in the response domain \cite{liu2018training}. For image-like free-form structures, generative models offer a more expressive output space. Liu \emph{et al.} used a generative framework for metasurfaces design \cite{liu2018generative}, Jiang \emph{et al.} generated free-form metagratings using GANs \cite{jiang2019free}, and So and Rho employed a conditional deep convolutional GAN for nanophotonic structures \cite{so2019designing}. Progressive growing was later used to increase resolution and stability \cite{wen2020robust}. Yeung \emph{et al.} trained a CNN to predict absorption spectra from resonator images and used SHAP explanations to expose learned shape--response relations \cite{yeung2020elucidating}. DeepAdjoint subsequently combined a learned generative model with adjoint optimization for multiobjective photonic design \cite{yeung2022deepadjoint}. A recent tandem-network study further demonstrated rapid metasurface parameter inference \cite{xu2023inverse}.

Current work increasingly treats inverse design as a distributional problem. Conditional diffusion models have generated metasurfaces from target scattering patterns \cite{hen2025inverse}, AdjointDiffusion incorporates adjoint sensitivity during sampling and explicitly considers fabrication constraints \cite{seo2026physics}, and MxDiffusion introduces a Maxwell law-guided two-stage diffusion strategy \cite{mondal2026mxdiffusion}. Mixture density networks have also been used to return multiple structural candidates for a single optical target \cite{cho2026probabilistic}. These methods directly address non-uniqueness, whereas the present study focuses on a compact deterministic reconstruction model and a controlled investigation of the spatial operator.

Table~\ref{tab:related_work} summarizes representative learning-based photonic inverse-design approaches according to their model family, structural output, and principal methodological contribution. Existing studies have primarily addressed the inverse problem through
forward model coupling, adversarial generation, adjoint refinement, or probabilistic sampling. However, comparatively limited attention has been given to the spatial operator used within image-generating decoders, even though this operator directly affects the reconstruction of boundaries, gaps, thin resonator elements, and component connectivity. This architectural gap motivates the investigation of adaptive spatial operators in the ensuing subsection.

\begin{table*}[t]
\centering
\caption{Representative learning-based approaches related to photonic inverse design. The studies differ in physical system, representation, and validation protocol; the table is contextual rather than a direct numerical comparison.}
\label{tab:related_work}
\small
\begin{tabularx}{\textwidth}{p{2.45cm}p{2.4cm}p{2.8cm}X}
\toprule
\textbf{Study} & \textbf{Model family} & \textbf{Inverse output} & \textbf{Main relevance} \\
\midrule
\cite{liu2018training} & Tandem neural network & Thin-film/design parameters & Uses a frozen forward model to alleviate inverse non-uniqueness. \\
\cite{liu2018generative} & Generative model & Metasurface geometry & Demonstrates image-generative inverse design from optical targets. \\
\cite{jiang2019free} & GAN & Free-form metagrating & Generates nonparametric diffractive structures. \\
\cite{so2019designing} & Conditional DCGAN & Nanophotonic image & Conditions a generative image model on desired optical properties. \\
\cite{wen2020robust} & Progressive GAN & High-resolution device image & Improves resolution and training stability for generative design. \\
DeepAdjoint \cite{yeung2022deepadjoint} & GAN + adjoint optimization & MIM metasurface & Combines data-driven global initialization with physics-based refinement. \\
\cite{xu2023inverse} & Deep tandem network & Metasurface parameters & Demonstrates fast target-spectrum-to-structure prediction. \\
\cite{hen2025inverse} & Conditional diffusion & Diffractive metasurface & Generates designs from spatial scattering targets and supports solver-guided sampling. \\
\cite{seo2026physics} & Physics-guided diffusion & Free-form photonic mask & Injects adjoint gradients and fabrication constraints into sampling. \\
\cite{mondal2026mxdiffusion} & Maxwell-guided diffusion & Photonic metasurface & Uses a physics-aware two-stage diffusion process. \\
\cite{cho2026probabilistic} & Mixture density network & Multiple MIM parameter sets & Explicitly models conditional non-uniqueness. \\
\textbf{This work} & \textbf{Deformable CNN + LSGAN} & \textbf{$64\times64$ MIM mask} & \textbf{Isolates adaptive spatial sampling and staged adversarial refinement.} \\
\bottomrule
\end{tabularx}
\end{table*}

\subsection{Adaptive Spatial Operators for Geometry Decoding}

As summarized in Table~\ref{tab:related_work}, contemporary nanophotonic inverse design studies have primarily addressed non-uniqueness and physical consistency by modifying the overall learning or optimization framework, including tandem networks, adversarial generation, adjoint refinement, and probabilistic modeling
\cite{liu2018training, liu2018generative, jiang2019free, so2019designing, wen2020robust, yeung2022deepadjoint, hen2025inverse, seo2026physics}. Comparatively less attention has been given to the spatial operator employed within image-generating decoders. This architectural choice is important in the present problem because, after the input spectrum is projected into a spatial latent representation, the decoder must progressively reconstruct thin resonator arms, sharp corners, internal apertures, narrow gaps, and disconnected components. Standard convolution applies a fixed Cartesian sampling pattern at every spatial location and may therefore be less flexible when reconstructing geometrically nonuniform structures
\cite{dai2017deformable, zhu2019deformable}.

Several adaptive operators provide different mechanisms for improving decoder flexibility. Dynamic convolution constructs an input-dependent combination of multiple kernels \cite{chen2020dynamic}, whereas ODConv extends this modulation across the spatial kernel, input channel, output channel, and kernel index dimensions \cite{li2022omni}. Involution generates location-specific spatial kernels that are shared across channels \cite{li2021involution}. These approaches adapt the filtering weights or kernels while retaining a regular spatial sampling neighborhood. Deformable convolution instead learns offsets for the sampling coordinates, allowing the receptive field itself to adapt to the spatial organization of the intermediate feature maps \cite{dai2017deformable, zhu2019deformable}.

This distinction motivates the use of deformable convolution in the proposed spectrum-to-geometry decoder. The input spectrum is first transformed into a $150\times4\times4$ latent tensor and then progressively expanded into a $64\times64$ resonator mask. During this reconstruction process, learned sampling offsets may provide greater flexibility in representing evolving boundaries, corners, gaps, and component layouts than fixed-grid convolution or kernel reweighting alone. Accordingly, deformable convolution is adopted as the proposed spatial operator and evaluated through a controlled comparison with plain convolution, Dynamic Conv, involution, and ODConv under the same generator architecture and two-stage training protocol.

\subsection{MIM Absorber Dataset}



The experiments use the open simulation dataset (\url{https://github.com/Raman-Lab-UCLA/Explainability_for_Photonics}) introduced by Yeung \emph{et al.} \cite{yeung2020elucidating}. The physical structure is a periodic MIM unit cell composed of a \SI{100}{\nano\meter} gold back reflector, a \SI{200}{\nano\meter} Al$_2$O$_3$ spacer, and a \SI{100}{\nano\meter} patterned gold resonator. The in-plane unit cell dimensions are $3.2~\mu\mathrm{m}\times3.2~\mu\mathrm{m}$, with periodic boundary conditions along the two in-plane axes. Three-dimensional FDTD simulations in Lumerical generated 10,000 structures and their absorption responses. The source study describes grayscale geometry images paired with 80 absorption samples over $4$--$12~\mu$m. These 80 positions values are treated as uniformly spaced wavelengths over $4$--$12~\mu$m for visualization and peak-error calculation. 

For the evaluation cycle, the 10,000 paired examples are divided into 80\% training, 10\% validation, and 10\% test partitions, corresponding to 8,000, 1,000, and 1,000 samples. Training batches are shuffled, whereas validation and test loaders are not.



\section{Proposed Method}
\label{sec:method}

\subsection{Inverse Reconstruction Task}

Let $\mathcal{D}=\{(\mathbf{s}_i,\mathbf{x}_i)\}_{i=1}^{N}$ denote paired absorption spectra and nanophotonic geometries, where $\mathbf{s}_i\in\mathbb{R}^{80}$ and $\mathbf{x}_i\in[0,1]^{1\times64\times64}$. The inverse model $G_{\theta}$ estimates the corresponding geometry as

$$
\widehat{\mathbf{x}}_i = G_{\theta}(\mathbf{s}_i)
$$

The model is trained using the geometry paired with each spectrum as the supervised target. Since the spectrum-to-geometry mapping can be non-unique, image domain agreement with the paired geometry does not necessarily imply that the alternative predictions are physically invalid. We therefore evaluate the reconstruction using both image domain metrics and spectral consistency
analysis through a separately trained forward model.

\subsection{Deformable Convolution-Based Inverse Generator}
\label{sec:generator}

\begin{figure*}[t]
\centering
\includegraphics[width=\textwidth]{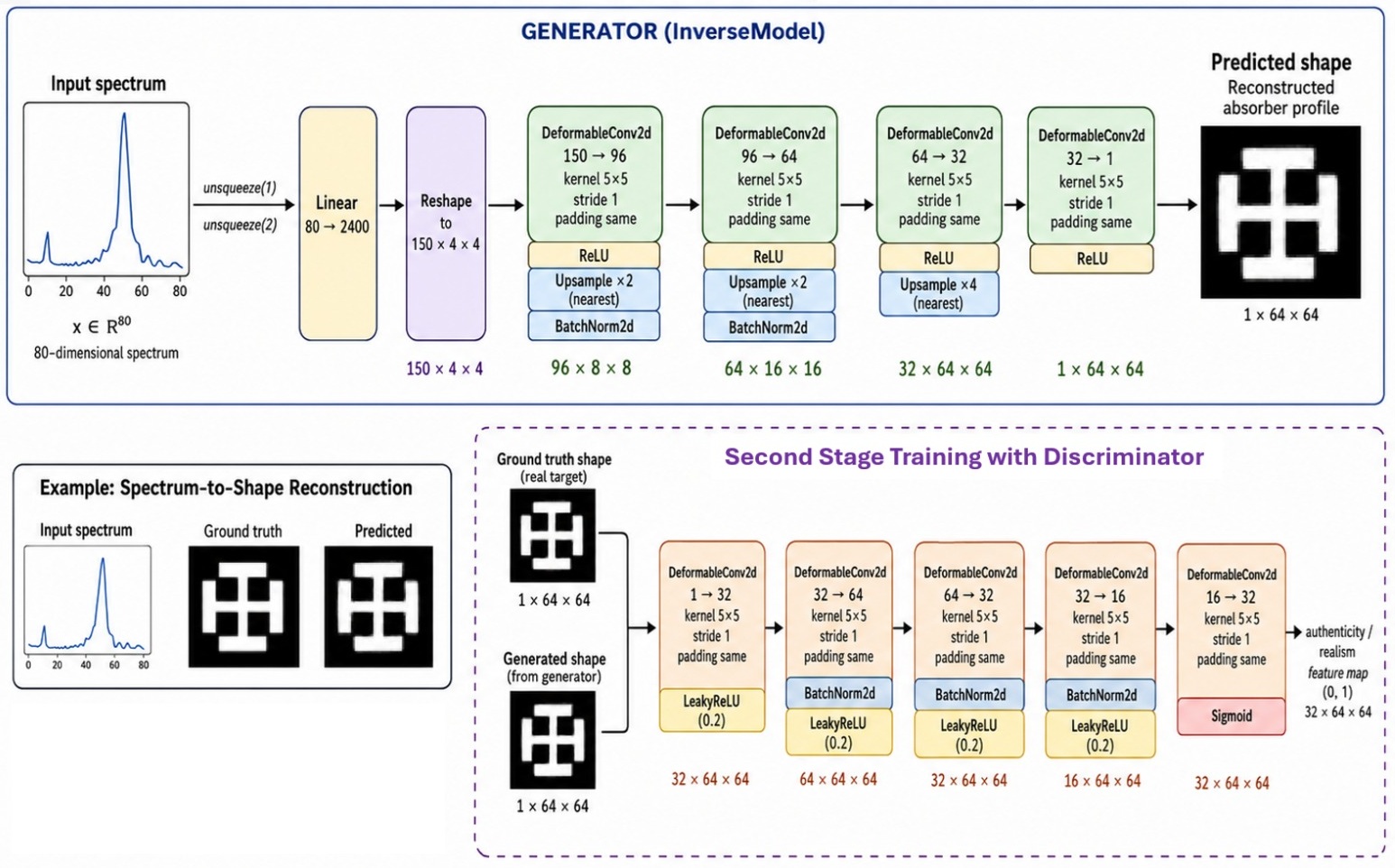}
\caption{Proposed spectrum-to-geometry inverse design framework. The target absorption spectrum is projected to a low-resolution spatial representation and progressively decoded using deformable convolutional layers. The generator is first optimized using supervised reconstruction and subsequently refined using a least-squares adversarial objective.}
\label{fig:architecture}
\end{figure*}

Fig.~\ref{fig:architecture} illustrates the proposed inverse generator. The 80-dimensional spectrum is first projected through a fully-connected layer and reshaped into a spatial latent representation,

$$
\mathbf{z}_0\in\mathbb{R}^{150\times4\times4}
$$

The latent representation is subsequently decoded through four spatial layers with channel dimensions,

$$
150\rightarrow96\rightarrow64\rightarrow32\rightarrow1
$$

while nearest-neighbor upsampling progressively increases the spatial resolution as,

$$
4\times4\rightarrow8\times8\rightarrow16\times16\rightarrow64\times64
$$

Each spatial layer employs a $5\times5$ kernel with unit stride and same padding. ReLU activations are used throughout the generator, while batch normalization is applied after the first two upsampling operations. In the proposed model, all four spatial layers are implemented using deformable convolution. For the architectural ablation study, these layers are replaced
by plain convolution, involution, Dynamic Conv, and ODConv while keeping the remaining decoder architecture unchanged.

\begin{table}[t]
\centering
\caption{Architecture of the inverse generator. The proposed model uses deformable convolution as the spatial operator.}
\label{tab:generator_architecture}
\scriptsize
\begin{tabular}{llll}
\toprule
\textbf{Step} & \textbf{Operation} & \textbf{Configuration} & \textbf{Output} \\
\midrule
Input & Spectrum & -- & $80$ \\
1 & Linear + reshape & $80\rightarrow2400$ & $150\times4\times4$ \\
2 & Spatial operator + ReLU & $150\rightarrow96$ & $96\times4\times4$ \\
3 & Upsample + BN & $\times2$ & $96\times8\times8$ \\
4 & Spatial operator + ReLU & $96\rightarrow64$ & $64\times8\times8$ \\
5 & Upsample + BN & $\times2$ & $64\times16\times16$ \\
6 & Spatial operator + ReLU & $64\rightarrow32$ & $32\times16\times16$ \\
7 & Upsample & $\times4$ & $32\times64\times64$ \\
8 & Spatial operator + ReLU & $32\rightarrow1$ & $1\times64\times64$ \\
\bottomrule
\end{tabular}
\end{table}

\subsection{Adaptive Spatial Sampling}
\label{sec:deformable}

A conventional convolution samples features from a fixed spatial grid. For a sampling set $\mathcal{R}=\{\mathbf{p}_k\}_{k=1}^{K}$, its output at location $\mathbf{p}_0$ is

$$
\mathbf{y}(\mathbf{p}_0) = \sum_{k=1}^{K} \mathbf{w}_k \mathbf{x}(\mathbf{p}_0+\mathbf{p}_k)
$$

Deformable convolution instead learns a displacement $\Delta\mathbf{p}_k$ for each sampling position. The implementation used in this work additionally learns a modulation coefficient $m_k$, yielding \cite{dai2017deformable, zhu2019deformable}

\begin{equation}
\mathbf{y}(\mathbf{p}_0) = \sum_{k=1}^{K} \mathbf{w}_k\, \mathbf{x} \left( \mathbf{p}_0+\mathbf{p}_k+\Delta\mathbf{p}_k \right) m_k 
\label{eq:deform}
\end{equation}

For the $5\times5$ kernels used in our model, $K=25$. Thus, each deformable layer predicts 50 offset values and 25 modulation coefficients at every spatial location. Fractional sampling positions are evaluated using bilinear interpolation.

The adaptive sampling mechanism is particularly suitable for the present inverse problem because the generated structures contain spatially varying features such as thin arms, corners, openings, and curved or irregular boundaries that may not be optimally represented by a fixed convolutional grid.

\subsection{Dense Discriminator}

During adversarial refinement, the discriminator $D_{\phi}$ receives either a reference geometry or a generated geometry. It consists of five $5\times5$ deformable convolutional layers with channel transitions as follows,

$$
1\rightarrow32\rightarrow64\rightarrow32\rightarrow16\rightarrow32
$$

Leaky ReLU activations are used after the first four layers, with batch normalization applied to the intermediate layers. A sigmoid activation is applied to the final dense authenticity map. The discriminator therefore provides spatially distributed adversarial feedback without flattening the geometry representation.

\begin{table}[!htb]
\centering
\caption{Architecture of the discriminator used during adversarial refinement.}
\label{tab:discriminator_architecture}
\scriptsize
\begin{tabular}{llll}
\toprule
\textbf{Layer} & \textbf{Channels} & \textbf{Kernel} &
\textbf{Post-operation} \\
\midrule
1 & $1\rightarrow32$ & $5\times5$ & LeakyReLU(0.2) \\
2 & $32\rightarrow64$ & $5\times5$ & BN + LeakyReLU(0.2) \\
3 & $64\rightarrow32$ & $5\times5$ & BN + LeakyReLU(0.2) \\
4 & $32\rightarrow16$ & $5\times5$ & BN + LeakyReLU(0.2) \\
5 & $16\rightarrow32$ & $5\times5$ & Sigmoid \\
\bottomrule
\end{tabular}
\end{table}

\subsection{Two-Stage Training}
\label{sec:training}

The proposed model is optimized in two stages. The first stage establishes the global spectrum-to-geometry mapping using supervised reconstruction. The second stage initializes the generator from the best supervised checkpoint and introduces adversarial learning to refine local structural details.

\subsubsection{Stage 1: Supervised Reconstruction}

The generator is first trained using pixel-wise mean-squared error,

\begin{equation}
\mathcal{L}_{\mathrm{rec}} = \mathbb{E}_{(\mathbf{s},\mathbf{x})} \left[ \left\|
G_{\theta}(\mathbf{s})-\mathbf{x} \right\|_2^2 \right]
\label{eq:recon}
\end{equation}

This stage allows the model to learn the overall topology and spatial layout of the resonator before introducing the adversarial objective.

\subsubsection{Stage 2: Least-Squares Adversarial Refinement}

The best generator obtained in Stage~1 initializes the second training stage. The discriminator is trained using a least-squares adversarial objective \cite{mao2017least},

\begin{equation}
\mathcal{L}_{D} = \frac{1}{2} \mathbb{E}_{\mathbf{x}} \left[ \|D_{\phi}(\mathbf{x})-\mathbf{1}\|_2^2 \right] + \frac{1}{2} \mathbb{E}_{\mathbf{s}} \left[ \|D_{\phi}(G_{\theta}(\mathbf{s}))\|_2^2 \right]
\label{eq:disc}
\end{equation}

The adversarial component used to update the generator is

\begin{equation}
\mathcal{L}_{\mathrm{adv}} = \mathbb{E}_{\mathbf{s}} \left[ \left\| D_{\phi}(G_{\theta}(\mathbf{s}))-\mathbf{1} \right\|_2^2 \right]
\label{eq:gadv}
\end{equation}

The final generator objective combines reconstruction fidelity and adversarial refinement as

\begin{equation}
\boxed{ \mathcal{L}_{G} = (1-\alpha_e)\mathcal{L}_{\mathrm{rec}} + \alpha_e\mathcal{L}_{\mathrm{adv}} }
\label{eq:final_loss}
\end{equation}

where $\alpha_e$ controls the contribution of adversarial learning. In the implemented schedule,

$$
\alpha_e= \begin{cases} 0, & e\leq50,\\ 0.1, & e>50. \end{cases}
$$

Thus, the second stage initially preserves pure reconstruction learning before activating adversarial refinement. This warm-start strategy prevents the discriminator from dominating the generator before a meaningful spectrum-to-geometry mapping has been established. The end-to-end training pipeline is presented in Algorithm~\ref{alg:training}.

\begin{algorithm}[!htbp]
\caption{Two-stage optimization of the inverse model}
\label{alg:training}
\KwIn{Training pairs $\mathcal{D}_{\mathrm{tr}}$ and validation set
$\mathcal{D}_{\mathrm{val}}$}
\KwOut{Refined generator $G_{\theta^\star}$}

Initialize $G_{\theta}$\;

\textbf{Stage 1: Supervised pretraining}\;

\For{each training epoch}{
    Update $G_{\theta}$ using $\mathcal{L}_{\mathrm{rec}}$\;
    Retain the checkpoint with the lowest validation loss\;
}

Initialize Stage~2 using the best generator checkpoint and initialize
$D_{\phi}$\;

\textbf{Stage 2: Adversarial refinement}\;

\For{each refinement epoch}{
    Set $\alpha_e=0$ during warm-up and $\alpha_e=0.1$ thereafter\;

    \If{$\alpha_e>0$}{
        Update $D_{\phi}$ using $\mathcal{L}_{D}$\;
    }

    Update $G_{\theta}$ using $\mathcal{L}_{G}$\;
    Retain the checkpoint with the lowest validation reconstruction loss\;
}

\Return{$G_{\theta^\star}$}\;
\end{algorithm}

\subsection{Deformable-Offset Analysis}
\label{sec:offset_analysis}

To investigate the spatial behavior learned by the proposed operator, we extract the deformable offsets from early, intermediate, and late decoder layers operating at $4\times4$, $16\times16$, and $64\times64$ resolution, respectively. Since the spectrum is first projected to a spatial latent representation, these offsets describe adaptive sampling in the decoder feature space rather than displacement along the original spectral vector.

For each layer, offset activity is summarized using the average displacement magnitude across the $K=25$ sampling points,

\begin{equation}
M_{\ell}(\mathbf{p}) = \frac{1}{K} \sum_{k=1}^{K} \left\| \Delta\mathbf{p}_{\ell,k}(\mathbf{p})
\right\|_2
\label{eq:offset_magnitude}
\end{equation}

The resulting maps are compared with structural boundaries, corners, and narrow gaps. We additionally visualize the regular and deformed sampling grids at representative locations to examine how the receptive field adapts during reconstruction. Since the final deformable layer operates directly at $64\times64$ resolution, its offsets provide the most direct spatial
correspondence with the reconstructed geometry.

\subsection{Forward-Surrogate Spectral Consistency}
\label{sec:spectral_consistency}

To complement image domain reconstruction metrics, the predicted geometry is evaluated using an independently trained forward surrogate $F_{\psi}$,

$$
\mathbf{s} \xrightarrow{G_{\theta}} \widehat{\mathbf{x}} \xrightarrow{F_{\psi}} \widehat{\mathbf{s}}
$$

The reconstructed spectrum is therefore

$$
\widehat{\mathbf{s}} = F_{\psi}\left(G_{\theta}(\mathbf{s})\right)
$$

and is compared directly with the target spectrum $\mathbf{s}$. Spectral consistency is quantified using MSE, RMSE, $R^2$, dominant-peak wavelength error, and peak-amplitude error. This experiment evaluates whether the inverse-generated geometry preserves the requested optical response and should be interpreted as \emph{forward-surrogate spectral validation}, rather
than full-wave FDTD re-simulation. 

\section{Experimental Setup}
\label{sec:experiments}

\subsection{Evaluation Protocol and Metrics}
\label{sec:evaluation_metrics}

All experiments use the fixed 80/10/10 training, validation, and test partition described in Section~\ref{sec:related}. Each experiment is repeated for three independent runs, and the results are reported as mean$\pm$sample standard deviation across the runs. For a metric $m_r$
obtained from run $r$, these statistics are calculated as follows

$$
\bar{m}=\frac{1}{R}\sum_{r=1}^{R}m_r,
\qquad
\sigma_m= \sqrt{\frac{1}{R-1} \sum_{r=1}^{R}(m_r-\bar{m})^2},
\qquad R=3
$$

\paragraph{Image domain reconstruction.}
Continuous-valued reconstruction quality is evaluated using mean-squared error (MSE), peak signal-to-noise ratio (PSNR), and structural similarity index (SSIM) \cite{wang2004image}. Let $\mathbf{x}$ and $\widehat{\mathbf{x}}$ denote the reference and reconstructed images,
respectively, with $M=H\times W$ pixels. The MSE is calculated as,

$$
\mathrm{MSE} = \frac{1}{M} \sum_{p=1}^{M} \left(x_p-\widehat{x}_p\right)^2
$$

Since the images are normalized to $[0,1]$, PSNR is calculated as

$$
\mathrm{PSNR} = 10\log_{10} \left( \frac{1}{\mathrm{MSE}} \right)
$$

SSIM evaluates local luminance, contrast, and structural agreement and is computed in its standard form,

$$
\mathrm{SSIM}(\mathbf{x},\widehat{\mathbf{x}}) = \frac{ (2\mu_x\mu_{\hat{x}}+C_1) (2\sigma_{x\hat{x}}+C_2) }{ (\mu_x^2+\mu_{\hat{x}}^2+C_1) (\sigma_x^2+\sigma_{\hat{x}}^2+C_2)}
$$

where $\mu$, $\sigma^2$, and $\sigma_{x\hat{x}}$ denote the local means, variances, and covariance, respectively, and $C_1$ and $C_2$ are the standard numerical-stability constants. The reported SSIM is averaged over the image.

\paragraph{Region and topology agreement.}
A more detailed geometric evaluation is performed for the selected two-stage DeformConv model. Predictions and references are binarized using a threshold of 0.5, yielding masks $\widehat{\mathbf{B}}$ and $\mathbf{B}$. Region overlap is quantified using the Dice coefficient and intersection-over-union (IoU), 

$$
\mathrm{Dice} = \frac{ 2|\mathbf{B}\cap\widehat{\mathbf{B}}|}{
|\mathbf{B}|+|\widehat{\mathbf{B}}| },
\qquad
\mathrm{IoU} = \frac{ |\mathbf{B}\cap\widehat{\mathbf{B}}|}{
|\mathbf{B}\cup\widehat{\mathbf{B}}|}
$$

These metrics characterize the agreement of the occupied material regions but do not explicitly measure boundary localization or connectivity.

\paragraph{Boundary agreement.}
Let $\partial\mathbf{B}$ and $\partial\widehat{\mathbf{B}}$ denote the reference and predicted
boundaries, and let

$$
d(\mathbf{p},\mathcal{S}) = \min_{\mathbf{q}\in\mathcal{S}} \|\mathbf{p}-\mathbf{q}\|_2
$$

denote the Euclidean distance from a point $\mathbf{p}$ to a set $\mathcal{S}$. Boundary precision and recall are computed using a tolerance of $\tau=2$ pixels,

$$
P_b = \frac{ |\{\mathbf{p}\in\partial\widehat{\mathbf{B}}: d(\mathbf{p},\partial\mathbf{B})\leq\tau\}| }{
|\partial\widehat{\mathbf{B}}|}
$$

$$
R_b = \frac{|\{\mathbf{p}\in\partial\mathbf{B}: d(\mathbf{p},\partial\widehat{\mathbf{B}})\leq\tau\}|}{
|\partial\mathbf{B}|}
$$

and the boundary F-score is calculated as

$$
\mathrm{Boundary\:F-Score} = \frac{2P_bR_b}{P_b+R_b}
$$

Surface displacement is further characterized using the 95th-percentile Hausdorff distance (HD95). Defining the set of bidirectional surface distances as

$$
\mathcal{D}_{s} = \{d(\mathbf{p},\partial\widehat{\mathbf{B}}): \mathbf{p}\in\partial\mathbf{B}\}
\cup \{d(\mathbf{q},\partial\mathbf{B}): \mathbf{q}\in\partial\widehat{\mathbf{B}}\}
$$

HD95 is calculated as

$$
\mathrm{HD95} = \operatorname{percentile}_{95}(\mathcal{D}_{s})
$$

The average surface distance (ASD) summarizes the average bidirectional boundary displacement,

$$
\mathrm{ASD} = \frac{1}{2} \left[ \frac{1}{|\partial\mathbf{B}|} \sum_{\mathbf{p}\in\partial\mathbf{B}} d(\mathbf{p},\partial\widehat{\mathbf{B}}) +
\frac{1}{|\partial\widehat{\mathbf{B}}|} \sum_{\mathbf{q}\in\partial\widehat{\mathbf{B}}}
d(\mathbf{q},\partial\mathbf{B}) \right]
$$

\paragraph{Connectivity and topology.}
Finally, structural connectivity is evaluated using connected-component count error and topology validity. If $C(\mathbf{B})$ denotes the number of 8-connected foreground components, the component-count error is 

$$
E_{\mathrm{CC}} = \left| C(\widehat{\mathbf{B}}) - C(\mathbf{B}) \right|
$$

Topology validity is determined from the Euler number $\chi(\cdot)$ under 8-connectivity. Over a test set of $N_t$ samples, the topology-valid fraction is

$$
T_{\mathrm{valid}} = \frac{1}{N_t} \sum_{i=1}^{N_t} \mathbb{I} \left[\chi(\widehat{\mathbf{B}}_i)
= \chi(\mathbf{B}_i) \right]
$$

where $\mathbb{I}[\cdot]$ is the indicator function. This metric is more sensitive than pixel-wise overlap to structural errors such as merged components, spurious islands, or incorrectly closed apertures.

\subsection{Operator and Training-Strategy Ablations}

To isolate the effect of the spatial operator, the same generator and discriminator architectures are retained throughout the ablation study. Only the spatial operator is replaced. Five alternatives are considered: plain convolution, deformable convolution, involution
\cite{li2021involution}, Dynamic Conv \cite{chen2020dynamic}, and ODConv \cite{li2022omni}. Channel dimensions, kernel sizes, upsampling operations, normalization, loss functions, and optimization settings remain unchanged.

Three optimization settings are evaluated. The \emph{supervised} setting trains the generator using only $\mathcal{L}_{\mathrm{rec}}$. The \emph{single-stage LSGAN} initializes the generator and discriminator from scratch and jointly optimizes reconstruction and adversarial objectives.
The proposed \emph{two-stage LSGAN} first learns the spectrum-to-geometry mapping through supervised reconstruction and then initializes adversarial refinement from the best supervised checkpoint. This comparison separates the contribution of the spatial operator from that of the staged optimization strategy.

All inverse models are implemented in PyTorch with a batch size of 32. The generator and discriminator are optimized using AdamW \cite{loshchilov2019decoupled}, with learning rates of $10^{-4}$ and $10^{-6}$, respectively. Stage~1 is trained for at most 100 epochs and Stage~2 for at most 500 epochs, with an early-stopping patience of 50 epochs based on validation pixel MSE. The adversarial term is activated after the warm-up period described in Section~\ref{sec:training}. No learning-rate scheduler is used.

\subsection{Forward-Surrogate Spectral Validation}

Image similarity alone cannot establish whether an inverse-generated geometry preserves the optical response that conditioned its generation. We therefore perform a spectrum--geometry--spectrum round-trip for the proposed model.

The forward validator is MobileNetV2-based regression model inspired by \cite{khan2026optimizing}. The model is independently trained using same training samples as inverse model to map a $64\times64$ binary geometry to an 80-point absorption spectrum. Its first convolution is adapted to a single input channel, and the classification head is replaced by a $1280\rightarrow512\rightarrow80$ regression head followed by a lightweight one-dimensional spectral refinement module. The model is trained using MSE and Adam with learning rate $10^{-3}$.

For every test spectrum $\mathbf{s}$, the inverse prediction is thresholded at 0.5 and evaluated using the frozen forward surrogate $F$:

$$
\mathbf{s} \rightarrow G(\mathbf{s}) \rightarrow \widehat{\mathbf{x}}_b \rightarrow F(\widehat{\mathbf{x}}_b)
$$

The input spectrum is compared with the corresponding forward-predicted spectrum using spectral MSE, RMSE, $R^2$, PSNR, dominant-peak wavelength error, and dominant-peak amplitude error. The peak location is evaluated on the uniformly represented 80-point wavelength grid spanning
$4$--$12~\mu$m.

This experiment is used as a scalable measure of \emph{forward-surrogate spectral consistency}; it is not treated as a substitute for full-wave electromagnetic simulation.

\subsection{Deformable-Offset Analysis}
\label{sec:offset_protocol}

To examine how adaptive sampling evolves during geometry reconstruction, the learned offsets and modulation coefficients are extracted from three representative DeformConv layers operating at $4\times4$, $16\times16$, and $64\times64$ spatial resolutions. These layers respectively characterize the early, intermediate, and late stages of the decoder. Because the input spectrum
is first projected to a spatial latent tensor, the extracted offsets describe adaptive sampling in the decoder feature space rather than displacement along the spectral dimension.

For a $5\times5$ deformable kernel, each spatial location contains $K=25$ learned two-dimensional offsets. Following Eq.~\eqref{eq:offset_magnitude}, their local activity is summarized by the mean displacement magnitude $M_{\ell}(\mathbf{p})$. To permit comparison
across layers having different spatial resolutions, the magnitude is normalized by the corresponding feature-map size,

$$
\widetilde{M}_{\ell}(\mathbf{p}) = \frac{M_{\ell}(\mathbf{p})} {\max(H_{\ell},W_{\ell})}
$$

where $H_{\ell}$ and $W_{\ell}$ denote the spatial dimensions of layer $\ell$. The modulation coefficients are additionally used to obtain a weighted mean displacement field,

$$
\overline{\Delta\mathbf{p}}_{\ell}(\mathbf{p}) = \frac{ \sum_{k=1}^{K} m_{\ell,k}(\mathbf{p})
\Delta\mathbf{p}_{\ell,k}(\mathbf{p}) }{
\sum_{k=1}^{K}m_{\ell,k}(\mathbf{p})+\epsilon}
$$

which provides a compact representation of the dominant sampling direction. For qualitative interpretation, normalized offset-magnitude maps, modulation-weighted displacement fields, mean modulation maps, and selected regular-versus-deformed $5\times5$ sampling grids are visualized for each decoder scale.

To relate the learned sampling behavior to the reconstructed geometry, structural regions are derived from the paired target mask after binarization at 0.5. A boundary band is obtained from the morphological gradient between one-step dilated and eroded masks and is subsequently expanded by two dilation iterations. Corners are detected using the Harris corner response ($\sigma=1$), with a minimum peak separation of two pixels and a relative response threshold of 0.08, the detected locations are then expanded by three binary-dilation iterations. Narrow gaps and openings are identified from background pixels lying between foreground regions in the horizontal or vertical direction over distances of one to four pixels. Enclosed holes are also included, and the resulting gap mask is expanded by one dilation iteration. Interior and background regions are defined by excluding the corresponding boundary and gap regions.

For comparison with these $64\times64$ geometric masks, the normalized offset-magnitude maps from the $4\times4$ and $16\times16$ layers are bilinearly interpolated to the output resolution. The $64\times64$ late-layer map requires no spatial resampling. For a structural region $\mathcal{R}$, offset concentration is quantified using the enrichment ratio

$$
E_{\ell,\mathcal{R}} = \frac{ \operatorname{mean}_{\mathbf{p}\in\mathcal{R}} \widetilde{M}_{\ell}(\mathbf{p}) }{
\operatorname{mean}_{\mathbf{p}\notin\mathcal{R}} \widetilde{M}_{\ell}(\mathbf{p})}
$$

Thus, $E_{\ell,\mathcal{R}}>1$ indicates greater offset activity within the specified region than elsewhere in the image, whereas values below one indicate the opposite tendency.

The spatial association between offset activity and structural boundaries is further assessed using the Spearman correlation between $\widetilde{M}_{\ell}$ and the Euclidean distance to the nearest target boundary. Negative correlation indicates that larger offsets tend to occur
closer to boundaries. Finally, one-sided paired Wilcoxon signed-rank tests compare the mean offset magnitude within boundary, corner, and gap/opening regions with that of the far-background region. The complete analysis is repeated for the independently trained models obtained with three different seeds.

\section{Results and Discussion}
\label{sec:results}

\subsection{Performance of the Proposed Two-Stage Deformable Model}
\label{sec:proposed_results}

We first evaluate the complete proposed configuration, comprising the DeformConv generator and supervised-to-LSGAN two-stage optimization. Across three independent runs, the model achieves
$20.79\pm0.31$~dB PSNR, $0.8501\pm0.0082$ SSIM, and $0.01378\pm0.00073$ MSE. The relatively small run-to-run variation indicates that the reconstruction performance is consistent across independent model initializations.

\begin{figure}[!htbp]
    \centering
    \includegraphics[width=0.95\linewidth]{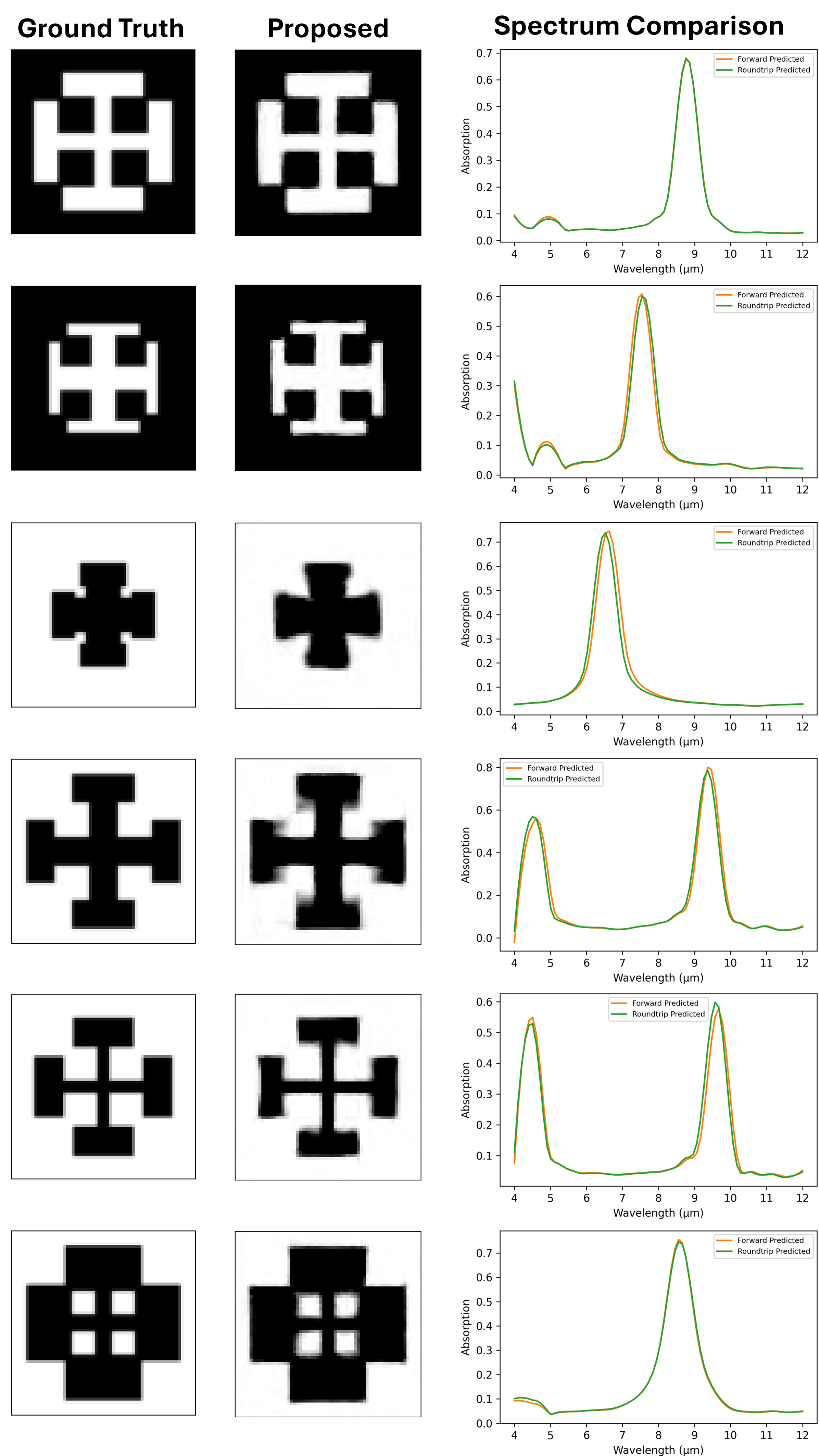}
    \caption{Representative forward-surrogate validation examples for the proposed two-stage DeformConv model. The target absorption spectrum is compared with the response predicted by the forward surrogate from the inverse model-generated geometry.}
    \label{fig:roundtrip_examples}
\end{figure}

These image domain results, presented in Fig.~\ref{fig:roundtrip_examples}, show that the model recovers the paired resonator geometry with good overall fidelity. However, pixel-wise metrics alone are insufficient for this problem. Small errors around an aperture, thin arm, or connecting region may have only a modest effect on PSNR or SSIM while substantially changing the geometry or its optical response. We therefore characterize the proposed model further through structural, spectral, and deformable-offset analyses.

\subsubsection{Geometric Fidelity of the Proposed Model}
\label{sec:geometric_fidelity}

The continuous-valued reconstruction metrics are complemented by a geometry-focused evaluation of the predictions. Table~\ref{tab:proposed_topology} summarizes region overlap, boundary localization, component connectivity, and topology preservation after thresholding the
predicted and reference masks at 0.5.

\begin{table}[]
\centering
\caption{Binary geometric evaluation of the proposed two-stage DeformConv model. Predictions and references are thresholded at 0.5. Boundary distances are reported in pixels. Values are reported for three independent trials and as mean$\pm$sample standard deviation.}
\label{tab:proposed_topology}
\scriptsize
\resizebox{\linewidth}{!}{
\begin{tabular}{lcccc}
\toprule
\textbf{Metric} &
\textbf{Trial-1} &
\textbf{Trial-2} &
\textbf{Trial-3} &
\textbf{Mean$\pm$SD} \\
\midrule
Dice $\uparrow$ &
0.9649 & 0.9625 & 0.9595 &
$0.9623\pm0.0027$ \\

IoU $\uparrow$ &
0.9380 & 0.9340 & 0.9305 &
$0.9342\pm0.0038$ \\

Boundary F-score $\uparrow$ &
0.9576 & 0.9523 & 0.9551 &
$0.9550\pm0.0027$ \\

HD95 $\downarrow$ &
1.8737 & 1.9971 & 1.7796 &
$1.883\pm0.109$ \\

ASD $\downarrow$ &
0.3340 & 0.3453 & 0.3795 &
$0.353\pm0.024$ \\

Connected-component error $\downarrow$ &
0.3382 & 0.3418 & 0.3055 &
$0.3285\pm0.0200$ \\

Topology-valid fraction $\uparrow$ &
0.7400 & 0.7300 & 0.7682 &
$0.7461\pm0.0198$ \\
\bottomrule
\end{tabular}}
\end{table}

The proposed model achieves a Dice coefficient of $0.9623\pm0.0027$ and an IoU of $0.9342\pm0.0038$, indicating strong agreement between the reconstructed and reference material regions. The boundary F-score of $0.9550\pm0.0027$ further shows that this agreement is not limited to the interior of the structures, but extends to their contours. Consistent with this observation, HD95 remains below two pixels ($1.883\pm0.109$), while the ASD is only
$0.353\pm0.024$ pixels. Together, these results indicate that the principal resonator elements and their boundaries are reconstructed with high spatial accuracy.


Topology preservation is more challenging. The mean topology-valid fraction for the three trials is $0.7461\pm0.0198$, while the mean connected-component count error is $0.3285\pm0.0200$. Thus, approximately three quarters of the reconstructed masks preserve the Euler topology of their paired references. The remaining cases can include small bridges, disconnected islands, merged components, or incorrectly closed openings. Such errors may occupy only a few pixels and therefore have limited influence on PSNR, SSIM, and Dice, while still changing the structural connectivity of the resonator. This distinction shows the value of evaluating topology in addition to conventional image-similarity measures.

\subsubsection{Spectral Consistency of the Reconstructed Designs}
\label{sec:spectral_results}

The preceding analysis, presented in Table~\ref{tab:proposed_topology} establishes similarity to the paired reference geometry. In an inverse-design setting, however, an equally important question is whether the reconstructed structure retains the optical response that originally conditioned the generator. We therefore evaluate the predictions using the independently trained forward surrogate described in Section~\ref{sec:experiments}.


\begin{figure*}[!htb]
    \centering
    \includegraphics[width=0.92\textwidth]{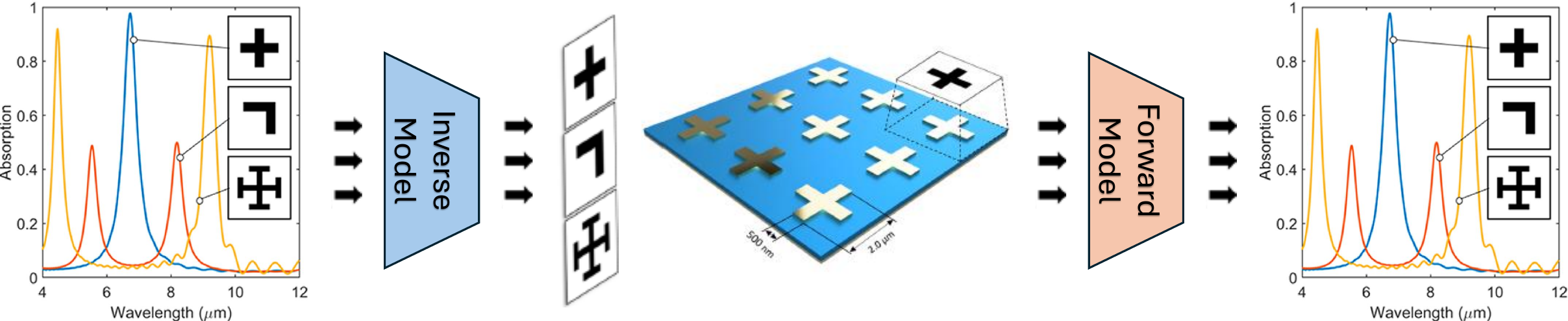}
    \caption{Forward-surrogate spectral validation of the proposed model. The target absorption spectrum is mapped to a geometry by the inverse model, and evaluated using the frozen forward surrogate. Spectral metrics compare the original target response with the response predicted for the reconstructed geometry.}
    \label{fig:roundtrip_workflow}
\end{figure*}

\begin{table*}[!htb]
\centering
\caption{Forward-surrogate spectral consistency of the geometry predictions from the proposed two-stage DeformConv model. Global spectral metrics are pooled over the test set, while dominant-peak errors are averaged over individual samples.}
\label{tab:roundtrip}
\scriptsize
\begin{tabular}{lcccccc}
\toprule
\textbf{Run} &
\textbf{MSE $\downarrow$} &
\textbf{RMSE $\downarrow$} &
$\mathbf{R^2\uparrow}$ &
\textbf{PSNR (dB) $\uparrow$} &
\textbf{Peak amp. error $\downarrow$} &
\textbf{Peak $\lambda$ error ($\mu$m) $\downarrow$} \\
\midrule


Trial-1 & 0.0068 & 0.0819 & 0.7851 & 21.677 & 0.2343 & 0.4251 \\

Trial-2 & 0.0065 & 0.0802 & 0.7942 & 21.865 & 0.2145 & 0.3921 \\

Trial-3 & 0.0064 & 0.0795 & 0.7977 & 21.940 & 0.2251 & 0.4385 \\

\midrule

Mean$\pm$SD &
$0.0066 \pm 0.0002$ &
$0.0805 \pm 0.0012$ &
$0.7923 \pm 0.0065$ &
$21.83 \pm 0.14$ &
$0.2246 \pm 0.0099$ &
$0.4186 \pm 0.0239$ \\
\bottomrule
\end{tabular}
\end{table*}

The reconstructed designs achieve a spectral RMSE of $0.0805\pm0.0012$ and an $R^2$ of $0.7923\pm0.0065$ averaged across the three runs. Mean spectral PSNR reaches $21.83\pm0.14$~dB, with little variation between the independently trained inverse models. These results indicate that a substantial portion of the target spectral response is retained after spectrum-to-geometry reconstruction.

Resonance-level errors provide a more demanding view of the same result. The dominant absorption peak is displaced by $0.4186\pm0.0239~\mu$m on average, while the corresponding peak-amplitude error is $0.2246\pm0.0099$. Thus, the round-trip is not exact. Importantly, the strong geometric overlap observed in Table~\ref{tab:proposed_topology} does not translate directly into equally high spectral agreement. This is reasonable because local geometric variations, for example, changes in gap width, arm length, or connectivity, can alter resonant behavior even when the overall mask remains visually similar.

\subsubsection{Analysis of Learned Deformable Offsets}
\label{sec:offset_results}

\begin{figure*}[!htb]
    \centering
    \includegraphics[width=\textwidth]{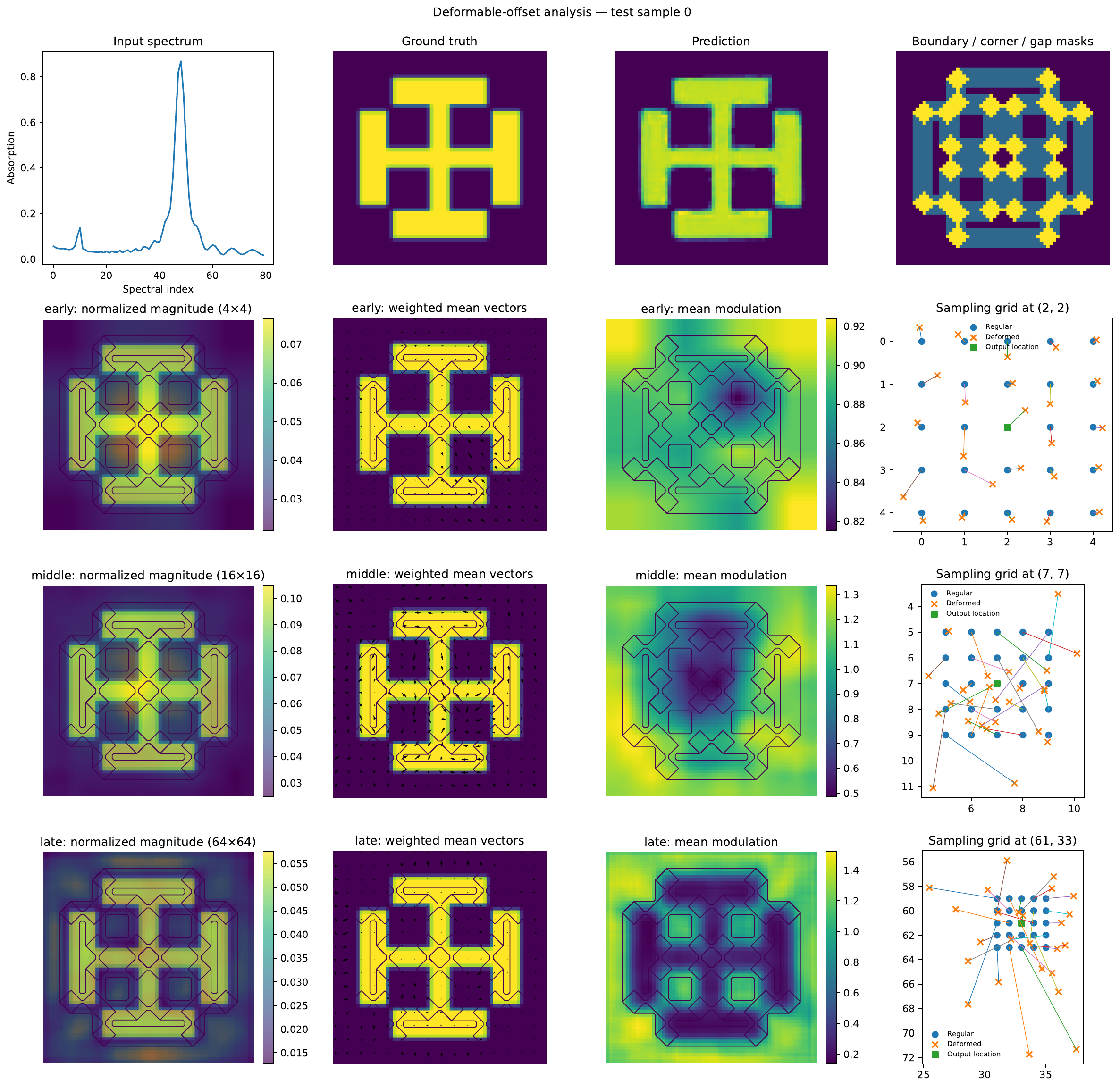}
    \caption{Representative multiscale analysis of the learned deformable offsets. The first row shows the input spectrum, reference geometry, reconstructed geometry, and structural-region masks. Subsequent rows show normalized offset magnitude, modulation-weighted displacement
    vectors, mean modulation, and a representative displaced sampling grid for the early $4\times4$, intermediate $16\times16$, and late $64\times64$ decoder layers.}
    \label{fig:offset_multiscale}
\end{figure*}

We next examine whether the learned sampling offsets provide insight into how deformable convolution contributes to the spectrum-to-geometry decoding. Fig.~\ref{fig:offset_multiscale} visualizes the normalized offset magnitudes, modulation-weighted displacement fields, modulation maps, and representative displaced sampling grids at the early $4\times4$, intermediate $16\times16$, and late $64\times64$ decoder stages.

The displacement patterns are clearly spatially non-uniform, indicating that the deformable layers learn location-dependent sampling rather than reducing to a uniform translation of the regular convolution grid. More importantly, the quantitative analysis in Table~\ref{tab:offset_summary} reveals that this behavior changes systematically with decoder depth.

\begin{table*}[t]
\centering
\caption{Multiscale deformable-offset statistics for the proposed two-stage model. Values are mean$\pm$sample standard deviation across trials 1--3. Enrichment greater than one indicates higher normalized offset magnitude within the specified structural region than outside that
region. The final column reports the Spearman correlation ($\boldsymbol{\rho}$) between normalized offset magnitude and distance from the nearest target boundary.}
\label{tab:offset_summary}
\footnotesize
\begin{tabular}{lcccccc}
\toprule
\textbf{Layer} &
\textbf{Grid} &
\makecell{\textbf{Mean normalized}\\\textbf{offset}} &
\makecell{\textbf{Boundary}\\\textbf{enrichment}} &
\makecell{\textbf{Corner}\\\textbf{enrichment}} &
\makecell{\textbf{Gap/opening}\\\textbf{enrichment}} &
\makecell{$\boldsymbol{\rho}$ \textbf{ with}\\\textbf{boundary distance}} \\
\midrule
Early &
$4\times4$ &
$0.0588\pm0.0019$ &
$1.188\pm0.004$ &
$1.238\pm0.003$ &
$1.628\pm0.017$ &
$-0.298\pm0.002$ \\

Middle &
$16\times16$ &
$0.0600\pm0.0010$ &
$1.049\pm0.026$ &
$1.064\pm0.034$ &
$1.204\pm0.052$ &
$-0.149\pm0.028$ \\

Late &
$64\times64$ &
$0.0264\pm0.0009$ &
$0.784\pm0.017$ &
$0.699\pm0.016$ &
$0.709\pm0.029$ &
$\phantom{-}0.266\pm0.031$ \\
\bottomrule
\end{tabular}
\end{table*}

At the early $4\times4$ stage, the learned offsets show a pronounced association with structural regions. Boundary and corner enrichment are $1.188\pm0.004$ and $1.238\pm0.003$, respectively, while the strongest concentration occurs around gaps and openings, with an enrichment ratio of
$1.628\pm0.017$. The negative correlation between offset magnitude and distance from the nearest boundary ($\rho=-0.298\pm0.002$) independently supports the same tendency.

The intermediate $16\times16$ layer retains this behavior, although with reduced strength. Boundary and corner enrichment remain slightly above unity, while gap/opening enrichment is $1.204\pm0.052$. Its negative boundary-distance correlation ($\rho=-0.149\pm0.028$) also indicates that larger displacements remain preferentially associated with locations closer
to structural boundaries. For both the early and intermediate stages, one-sided Wilcoxon tests show significantly larger offset magnitudes in the boundary, corner, and gap/opening regions than in the far-background region ($p<8.2\times10^{-57}$).

The late $64\times64$ layer exhibits a different pattern. Its mean normalized offset magnitude decreases to $0.0264\pm0.0009$, less than half that observed at the earlier scales. Boundary, corner, and gap/opening enrichment all fall below one, and the boundary-distance correlation becomes positive ($\rho=0.266\pm0.031$). The final deformable layer therefore does not behave primarily as a direct boundary-tracing mechanism.

Taken together, the results suggest a scale-dependent role for deformable sampling. The strongest geometry-associated displacement occurs while the decoder is transforming the coarse latent representation into the global and intermediate arrangement of resonator components. Once this spatial organization has been established, the final layer appears to require smaller and more distributed adjustments. This interpretation is more consistent with the observed data than attributing the DeformConv advantage solely to late-stage edge refinement. The offset analysis nevertheless remains associative, it reveals where adaptive sampling occurs, but does not
by itself establish the causal contribution of individual offsets to the final reconstruction accuracy.

\subsection{Effect of Two-Stage Optimization}
\label{sec:training_ablation_results}

Having established the performance and behavior of the complete proposed model, we next isolate the contribution of its two-stage optimization strategy. To avoid confounding the training analysis with spatial-operator choice, all three configurations in Table~\ref{tab:stage_ablation} use the same DeformConv generator.

\begin{table}[t]
\centering
\caption{Training-strategy ablation using the DeformConv generator. Values are mean$\pm$sample standard deviation over three independent runs.}
\label{tab:stage_ablation}
\scriptsize
\resizebox{\linewidth}{!}{
\begin{tabular}{lccc}
\toprule
\textbf{Training strategy} &
\textbf{PSNR (dB)} &
\textbf{SSIM} &
\textbf{MSE} \\
\midrule



Supervised &
$20.34\pm0.22$ &
$0.8373\pm0.0064$ &
$0.00924\pm0.00161$ \\

Single-stage LSGAN &
$17.12\pm0.19$ &
$0.7394\pm0.0061$ &
$0.01941\pm0.00035$ \\

\textbf{Proposed two-stage LSGAN} &
$\mathbf{20.79\pm0.31}$ &
$\mathbf{0.8501\pm0.0082}$ &
$\mathbf{0.00834\pm0.00073}$ \\
\bottomrule
\end{tabular}}
\end{table}

Supervised reconstruction alone already provides a strong baseline, reaching $20.34\pm0.22$~dB PSNR and $0.8373\pm0.0064$ SSIM. In contrast, introducing adversarial optimization directly from random initialization substantially degrades performance, reducing PSNR to $17.12\pm0.19$~dB and SSIM to $0.7394\pm0.0061$. This result indicates that the adversarial objective alone does not provide a sufficiently stable signal for learning the spectrum-to-geometry correspondence from an uninformative initialization.

Initializing Stage~2 from the supervised solution changes this behavior. The proposed two-stage strategy improves PSNR by 0.45~dB and SSIM by 0.0128 relative to supervised reconstruction alone, while reducing MSE from $0.00924$ to $0.00834$, corresponding to a 9.74\% reduction. Relative to single-stage LSGAN training, the improvement is considerably larger, where PSNR increases by 3.67~dB and MSE decreases by 57.03\%.

These results support the central motivation for staged optimization. Supervised learning first establishes the global conditional mapping between the input spectrum and the resonator layout. Adversarial refinement is then introduced from an already meaningful geometry, allowing the discriminator to influence local structure without simultaneously forcing the generator
to discover the overall layout.

\begin{figure*}[t]
    \centering
    \includegraphics[width=\textwidth]{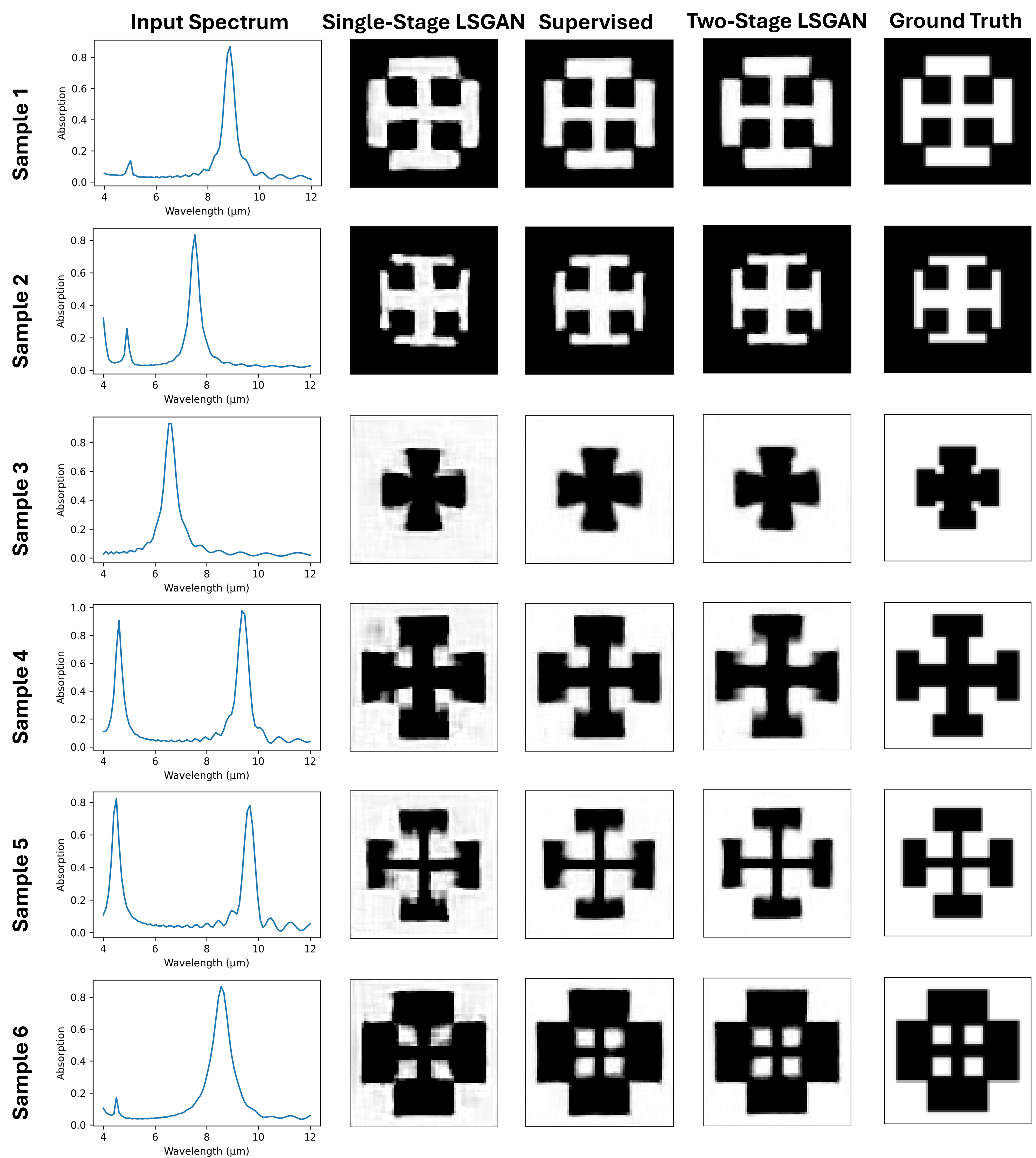}
    \caption{Representative reconstructions obtained using the same DeformConv generator under supervised training, single-stage LSGAN optimization, and the proposed two-stage strategy. Supervised initialization preserves the global spectrum-to-geometry correspondence,
    while subsequent adversarial refinement improves local structural definition.}
    \label{fig:qualitative_stages}
\end{figure*}

The qualitative examples in Fig.~\ref{fig:qualitative_stages} support the quantitative findings. Supervised predictions generally recover the dominant geometry but retain softer boundaries and less uniform structural widths. Single-stage adversarial optimization can introduce larger distortions, whereas the proposed two-stage model preserves the established layout while
producing cleaner boundaries and openings. Thus, the gain obtained from the second stage should be interpreted as refinement of a learned inverse mapping, rather than adversarial learning replacing supervised reconstruction.

\subsection{Comparison with Alternative Spatial Operators}
\label{sec:operator_results}

We finally assess whether the performance of the proposed model is specific to deformable sampling or can be reproduced by alternative adaptive spatial operators. Table~\ref{tab:main_results} compares plain convolution, DeformConv, involution, Dynamic Conv, and ODConv under the same surrounding decoder architecture and training protocol.

\begin{table*}[t]
\centering
\caption{Three-run comparison of spatial operators. Stage~1 denotes supervised reconstruction, while the two-stage configuration initializes adversarial refinement from the corresponding supervised model. Values are mean$\pm$sample standard deviation; bold indicates the highest mean PSNR and SSIM values.}
\label{tab:main_results}
\scriptsize
\begin{tabular}{lcccccc}
\toprule
\multirow{2}{*}{\textbf{Operator}} &
\multicolumn{2}{c}{\textbf{Supervised}} &
\multicolumn{2}{c}{\textbf{Two-stage LSGAN}} &
\multicolumn{2}{c}{\textbf{Gain}} \\
\cmidrule(lr){2-3}
\cmidrule(lr){4-5}
\cmidrule(lr){6-7}
&
\textbf{PSNR (dB)} &
\textbf{SSIM} &
\textbf{PSNR (dB)} &
\textbf{SSIM} &
$\Delta$\textbf{PSNR} &
$\Delta$\textbf{SSIM} \\
\midrule
Plain convolution &
$18.05\pm0.16$ &
$0.7444\pm0.0060$ &
$18.63\pm0.11$ &
$0.7670\pm0.0071$ &
$+0.58$ &
$+0.0226$ \\

Involution &
$14.54\pm0.04$ &
$0.5737\pm0.0111$ &
$16.13\pm0.86$ &
$0.6742\pm0.0505$ &
$+1.59$ &
$+0.1004$ \\

Dynamic convolution &
$13.94\pm0.70$ &
$0.5585\pm0.0566$ &
$14.27\pm0.70$ &
$0.5814\pm0.0507$ &
$+0.33$ &
$+0.0228$ \\

ODConv &
$17.94\pm0.08$ &
$0.7371\pm0.0065$ &
$19.17\pm0.17$ &
$0.7838\pm0.0038$ &
$+1.23$ &
$+0.0467$ \\

\textbf{Deformable convolution (proposed)} &
$\mathbf{20.34\pm0.22}$ &
$\mathbf{0.8373\pm0.0064}$ &
$\mathbf{20.79\pm0.31}$ &
$\mathbf{0.8501\pm0.0082}$ &
$+0.45$ &
$+0.0128$ \\
\bottomrule
\end{tabular}
\end{table*}

DeformConv is the strongest operator already in the supervised setting, where it reaches $20.34\pm0.22$~dB PSNR and $0.8373\pm0.0064$ SSIM. This is important because the advantage therefore precedes adversarial refinement and cannot be attributed solely to the second training stage.

After two-stage optimization, the proposed configuration reaches $20.79\pm0.31$~dB and $0.8501\pm0.0082$ SSIM. Relative to plain convolution, this corresponds to gains of 2.16~dB PSNR and 0.0831 SSIM. ODConv is the strongest alternative adaptive operator, reaching $19.17\pm0.17$~dB and $0.7838\pm0.0038$, but remains 1.62~dB and 0.0663 SSIM below DeformConv.
Involution and Dynamic Conv perform substantially worse under the same decoder and optimization settings.


The magnitude of the refinement gain varies across the operators. Involution and ODConv show larger absolute improvements after the second stage ($+1.59$ and $+1.23$~dB, respectively), but both start from substantially weaker supervised solutions. DeformConv improves by a smaller additional $0.45$~dB because its supervised model is already considerably stronger. This distinction is important as a large Stage~2 gain does not imply a better final inverse model. Instead, the results indicate that adversarial refinement can improve several spatial operators, while the quality of the underlying spectrum-to-geometry mapping remains strongly dependent on the operator itself.

The comparative results also suggest that not all forms of spatial adaptation are equally suitable for this reconstruction problem. Dynamic Conv and ODConv adapt convolutional kernels, while involution employs location-dependent spatial kernels. DeformConv differs in that it explicitly changes the spatial coordinates from which features are sampled. Given the thin arms, openings, irregular component layouts, and spatial transitions present in the resonator masks, adapting the sampling geometry appears to be particularly effective.

\begin{figure*}[t]
    \centering
    \includegraphics[width=0.98\textwidth]{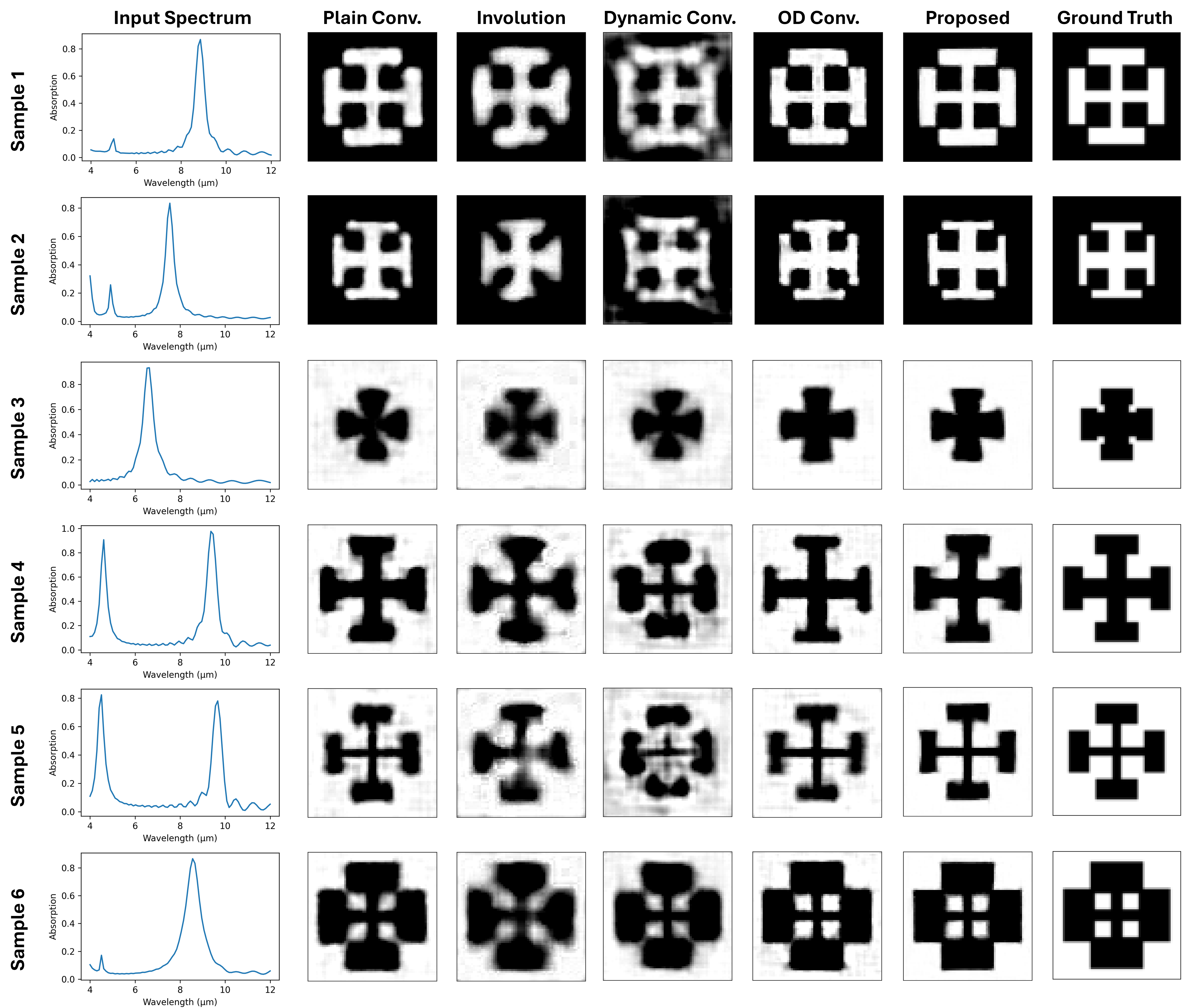}
    \caption{Representative two-stage reconstructions obtained with different spatial operators. Columns show the input spectrum, reference geometry, and predictions from plain convolution, involution, Dynamic Conv, ODConv, and the proposed DeformConv model.}
    \label{fig:qualitative}
\end{figure*}

The qualitative comparison in Fig.~\ref{fig:qualitative} is consistent with the aggregate metrics. Dynamic Conv predictions frequently exhibit diffused regions or background leakage, while involution tends to smooth corners and broaden narrow structures. Plain convolution generally recovers the dominant silhouette, and ODConv improves the reconstruction of major
components but can retain irregular widths or deformed apertures. The proposed DeformConv most consistently preserves separated components, narrow connecting elements, straight boundaries, and internal openings. Together with the offset analysis in Section~\ref{sec:offset_results}, these findings support adaptive sampling coordinates as the principal architectural distinction of the proposed decoder.

\subsection{Limitations and Validation Scope}
\label{sec:limitations}

The results should be interpreted within several limitations. First, spectral validation is performed using an independently trained forward surrogate rather than direct FDTD re-simulation. Although the surrogate enables test set-wide evaluation and the round-trip results show meaningful response preservation, generated geometries may differ from the distribution on which the forward model was trained. Consequently, the reported spectral metrics should be interpreted as surrogate-based consistency rather than direct electromagnetic verification. Full-wave simulation of representative generated designs would provide a stronger independent assessment.


Second, the proposed model remains deterministic and returns a single geometry for a given target spectrum. Because the inverse mapping is non-unique, this formulation does not explicitly explore alternative geometries that may realize comparable optical responses. Probabilistic or
diffusion-based formulations could complement the present approach when solution diversity is required \cite{hen2025inverse, seo2026physics, mondal2026mxdiffusion, cho2026probabilistic}.

Third, the deformable-offset analysis provides evidence that adaptive sampling is systematically associated with structural regions at the early and intermediate decoder stages, but the analysis is observational rather than causal. Suppressing, freezing, or randomizing offsets at individual layers would be required to quantify the causal contribution of each
sampling scale to reconstruction accuracy.

Finally, topology remains less reliable than region overlap. Although the proposed model achieves Dice above 0.96 and a boundary F-score above 0.95, the topology-valid fraction is approximately 0.75. Future work could therefore incorporate topology-aware or connectivity-preserving objectives to reduce small bridges, disconnected components, and incorrectly closed
openings. Fabrication-aware constraints, computational-efficiency analysis, and explicit multi-solution generation are outside the scope of the present study and are not used to support its claims.
\section{Conclusion}
\label{sec:conclusion}

This work investigated deformable spatial sampling for spectrum-to-geometry reconstruction of metal--insulator--metal nanophotonic resonators. The proposed two-stage framework first learns the global mapping through supervised reconstruction and then refines the generator using least-squares adversarial training. Across three independent runs, the DeformConv model achieved the best performance among the evaluated spatial operators, reaching $20.79\pm0.31$~dB PSNR and $0.8501\pm0.0082$ SSIM. Binary geometric evaluation further showed strong structural agreement, with a Dice score of $0.9623\pm0.0027$ and boundary F-score of $0.9550\pm0.0027$.

Forward-surrogate evaluation of the thresholded predictions yielded a spectral RMSE of $0.0805\pm0.0013$ and $R^2=0.7923\pm0.0065$, indicating meaningful preservation of the conditioning response. The deformable-offset analysis further revealed scale-dependent behavior, with stronger geometry-associated displacement at the coarse and intermediate decoder stages than at the final output resolution.

Overall, the results demonstrate that deformable spatial sampling combined with supervised initialization and adversarial refinement is effective for spectrum-conditioned geometry reconstruction. The generated structures should nevertheless be regarded as structurally faithful and surrogate-consistent candidate designs, while full-wave electromagnetic simulation remains necessary for independent physical verification.


\printcredits

\section*{Declaration of competing interest}
The authors declare that they have no known competing financial interests or personal relationships that could have appeared to influence the work reported in this paper. 

\section*{Data and code availability}
The original MIM simulation dataset is associated with Yeung \emph{et al.} \cite{yeung2020elucidating}. Public repository and trained-checkpoint are available at author's Github \url{https://github.com/eeshahid/nanophotonic-inverse-design} .

\section*{Acknowledgment}
The authors acknowledge support from the Saudi Data and AI Authority (SDAIA) and King Fahd University of Petroleum and Minerals (KFUPM) through the SDAIA-KFUPM Joint Research Center for Artificial Intelligence. Shujaat Khan would also like to acknowledge the support from KFUPM under Early Career Grant No EC241027.

\end{document}